\documentclass[conference]{IEEEtran}
\IEEEoverridecommandlockouts
\usepackage{cite}
\usepackage{amsmath,amssymb,amsfonts}
\usepackage{algorithmic}
\usepackage{graphicx}
\usepackage{textcomp}
\usepackage{xcolor}
\def\BibTeX{{\rm B\kern-.05em{\sc i\kern-.025em b}\kern-.08em
    T\kern-.1667em\lower.7ex\hbox{E}\kern-.125emX}}

\usepackage{commath}

\usepackage{booktabs}
\usepackage{multirow}
\usepackage[flushleft]{threeparttable}

\ifCLASSOPTIONcompsoc
\usepackage[caption=false,font=normalsize,labelfont=sf,textfont=sf]{subfig}
\else
\usepackage[caption=false,font=footnotesize]{subfig}
\fi

\begin{document}

\title{Performance Modelling of Deep Learning on Intel Many Integrated Core Architectures \thanks{This research has received funding from the Swedish Knowledge Foundation under Grant No. 20150088}
}

\author{
    \IEEEauthorblockN{Andre Viebke}
	\IEEEauthorblockA{
		\textit{Linnaeus University}\\
		V\"{a}xj\"{o}, Sweden \\
		av22cj@student.lnu.se}
	\and
	\IEEEauthorblockN{Sabri Pllana}
	\IEEEauthorblockA{
		\textit{Linnaeus University}\\
		V\"{a}xj\"{o}, Sweden \\
		sabri.pllana@lnu.se}
	\and
    \IEEEauthorblockN{Suejb Memeti}
	\IEEEauthorblockA{
		\textit{Link\"{o}ping University}\\
		Link\"{o}ping, Sweden \\
		suejb.memeti@liu.se}
	\and
	\IEEEauthorblockN{Joanna Kolodziej}
	\IEEEauthorblockA{
		\textit{Cracow University of Technology}\\
		Cracow, Poland \\
		jokoldziej@pk.edu.pl}
}


\IEEEspecialpapernotice{\footnotesize(Preprint, HPCS, \copyright 2019 IEEE)}

\maketitle

\begin{abstract}
Many complex problems, such as natural language processing or visual object detection, are solved using deep learning. However, efficient training of complex deep convolutional neural networks for large data sets is computationally demanding and requires parallel computing resources. 
In this paper, we present two parameterized performance models for estimation of execution time of training convolutional neural networks on the Intel many integrated core architecture. While for the first performance model we minimally use measurement techniques for parameter value estimation, in the second model we estimate more parameters based on measurements. We evaluate the prediction accuracy of performance models in the context of training three different convolutional neural network architectures on the Intel Xeon Phi. The achieved average performance prediction accuracy is about 15\% for the first model and 11\% for second model. 
\end{abstract}


\section{Introduction}
\label{sec:introduction}

Deep learning \cite{LeCun:2015} is modeled as artificial deep neural network that use many processing layers to learn complex functions with successful application in various domains including, self-driving cars \cite{Bojarski:2016}, object recognition \cite{Krizhevsky:2017}, natural language processing \cite{Collobert:2011}, speech recognition \cite{Hinton:2012}, language translation \cite{Sutskever:2014}, optimization of the Cloud \cite{Grzonka:2018} and parallel computing systems \cite{Memeti:2018}.

\begin{figure}[tb]
		\center
		\includegraphics[width=\linewidth]{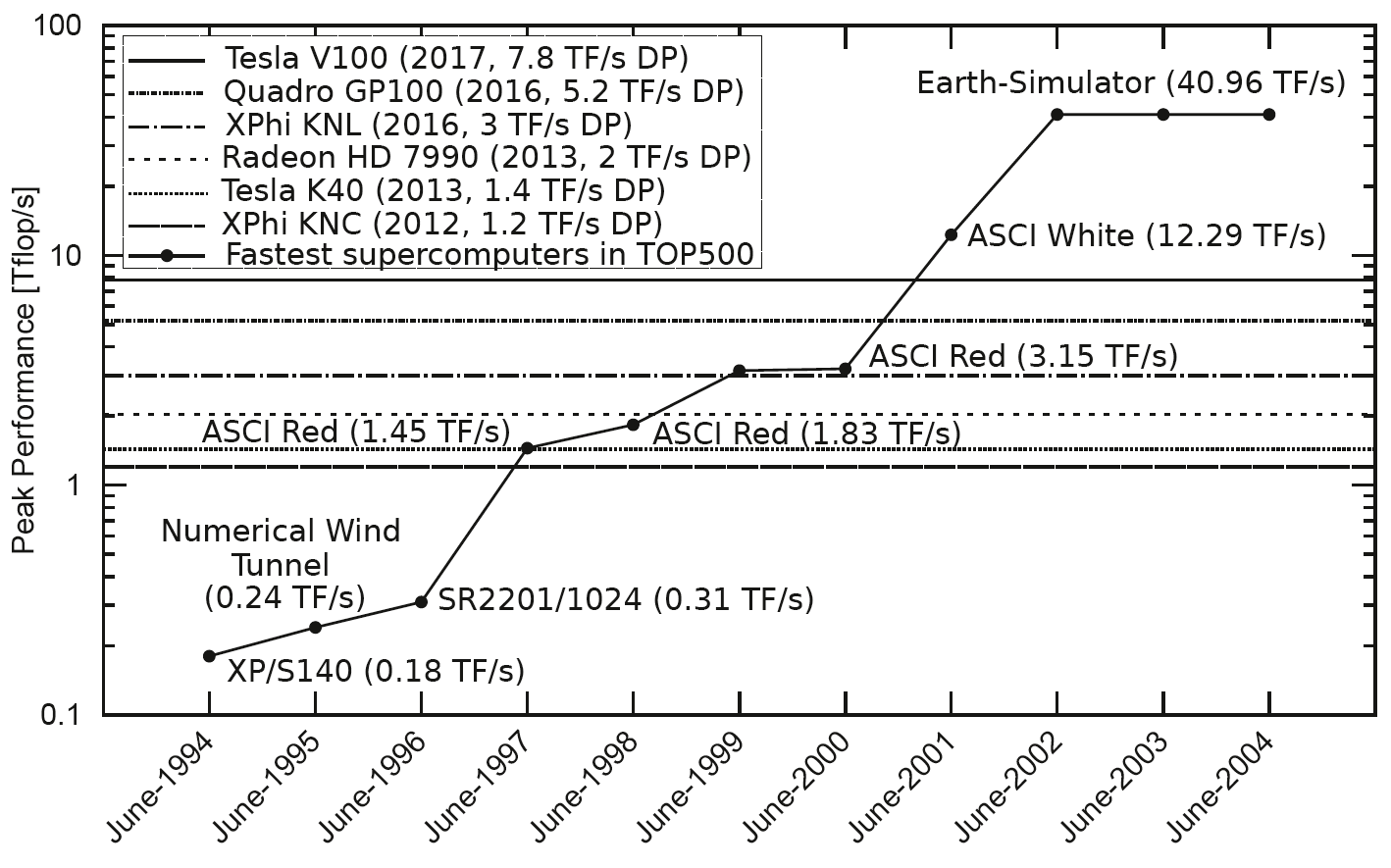}
		\caption{Performance of many-core processors and \#1 in the TOP500 list \cite{top500} of most powerful supercomputers. The Intel Xeon Phi KNL in 2016 offered similar performance like the supercomputer ASCI RED that was \#1 in June 2000.}
		\label{fig:accelerators}
	\end{figure}

Deep learning is becoming increasingly computational demanding in accordance with the trend of increasing volumes of available data \cite{Vitabile:2019} and complexity of deep neural networks. Therefore, many-core parallel computing systems \cite{Czarnul:2018,pllana:2017,Smari:2016,peppher:2011,Abraham:2015} are used to accelerate the learning process of  deep neural networks \cite{Hoefler:2018,Zlateski:2016}. Many-core processors, such as NVIDIA GPU or the Intel Xeon Phi, provide high performance and may be used to accelerate the process of deep learning. Figure \ref{fig:accelerators} depicts the performance of many-core processors compared to the fastest supercomputers in the world in the TOP500 list \cite{top500}. For instance, the peak performance of the Intel Xeon Phi Knights Corner (KNC) or the Tesla K40 is similar to the fastest supercomputer in the year 1997 that was ASCI Red with 1.45 Teraflop/s peak performance.

Related work has studied performance modeling of deep learning on distributed systems \cite{yan2015}, performance prediction of asynchronous stochastic gradient descent \cite{oyama2016}, performance modelling of various distributed deep learning frameworks (such as, Caffe-MPI or TensorFlow) \cite{shi:2018}, analytical models for predicting the usage of optimal resources of a GPU for deep learning \cite{song2017}. However, not much attention has been devoted to performance modeling of deep convolutional neural networks on the Intel many integrated core architecture. 

In this paper, we describe our approach for performance modeling of training convolutional neural networks on the Intel Xeon Phi many core processor. We develop two parameterized performance models based on the theoretical analysis of a code \cite{Dan:code} for training convolutional neural networks that we parallelized for the Intel Xeon Phi. Input variables of the performance models are the number of training or validation images, the number of test images, the number of network instances, the number of epochs, and the number of processing units. For the development of the first performance model, we minimally use measurements for estimating parameter values of the performance model; only memory contention is estimated using measurements. For the second performance model, we apply measurements for estimation of the sequential work, and the forward- and back-propagation. For evaluation, we use the MNIST [20] data-set of handwritten digits. The average deviation of predicted from the measured performance over all measured thread counts and various neural network architectures is about 15\% for the first model and 11\% for second model. Major contributions of this paper include,
\begin{itemize}
    \item development of two performance models for estimation of execution time of training convolutional neural networks on the Intel Xeon Phi,
    \item evaluation of prediction accuracy of performance models for various execution contexts and neural network architectures,
    \item model-driven performance evaluation for larger number of threads than the number of hardware threads of the Intel Xeon Phi under study. 
\end{itemize}

The rest of this paper is structured as follows. Section \ref{sec:nn} gives an overview of convolutional neural networks that are addressed in this paper. We describe Intel many integrated core architecture in Section \ref{sec:xphi}. Section \ref{sec:performance-model} describes our performance modelling approach. An empirical evaluation of the performance model is described in Section \ref{sec:evaluation}. We discuss the related work in Section \ref{sec:rw}. Section \ref{sec:summary} concludes this paper.

\section{Convolutional Neural Networks}
\label{sec:nn}
	
An artificial deep neural network is the underlying model used in deep learning \cite{LeCun:2015}. A Convolutional Neural Network (CNN) is a variant of a Deep Neural Network (DNN), which introduces two additional layer types: \emph{convolutional layers} and \emph{pooling layers}. The mammal visual processing system is hierarchical (deep) in nature. Higher level features are abstractions of lower level ones. For instance, to understand speech, waveforms are translated through several layers until reaching a linguistic level. A similar analogy can be drawn for images, where edges and corners are lower level abstractions translated into more spatial patterns on higher levels.  
	
The architecture of a DNN consists of multiple layers of neurons. Neurons are connected to each other through edges (weights). The network can simply be thought of as a weighted graph; a directed acyclic graph represents a feed-forward network. The depth and breadth of the network differs as may the layer types. Regardless of the depth, a network has at least one input and one output layer. A neuron has a set of incoming weights, which have corresponding outgoing edges attached to neurons in the previous layer. Also, a bias term is used at each layer as an intercept term. The goal of the learning process is to adjust the network weights and find a global minimum by reducing the overall error, i.e. the deviation between the predicted and the desired outcome of all the samples. The resulting weight parameters can thereafter be used to make predictions of unseen inputs \cite{ng2011ufldl}.

\begin{figure} [tb]
	\centering
    \subfloat[Small CNN: the first convolutional layer has 5 maps, 3380 neurons, uses a kernel size of 4x4, a map size of 26x26 and 85 weights.]
        {\includegraphics[width=0.8\linewidth]{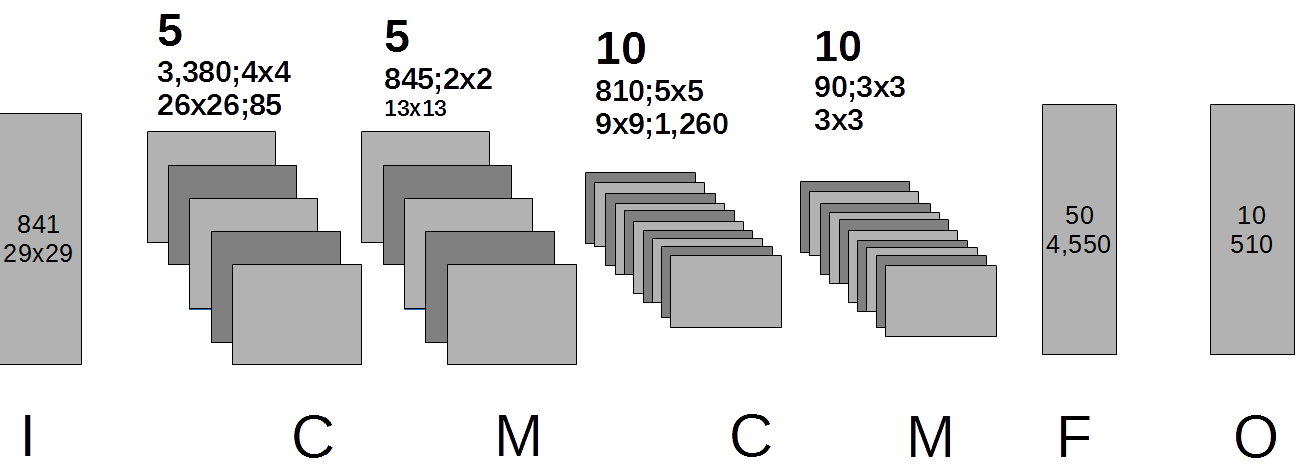}
    	\label{fig:cnn-small}}
	\hfill
    \subfloat[Medium CNN: the first convolutional layer has 20 maps, 13,520 neurons, uses a kernel size of 4x4, a map size of 26x26 and 340 weights.]
       {\includegraphics[width=0.9\linewidth]{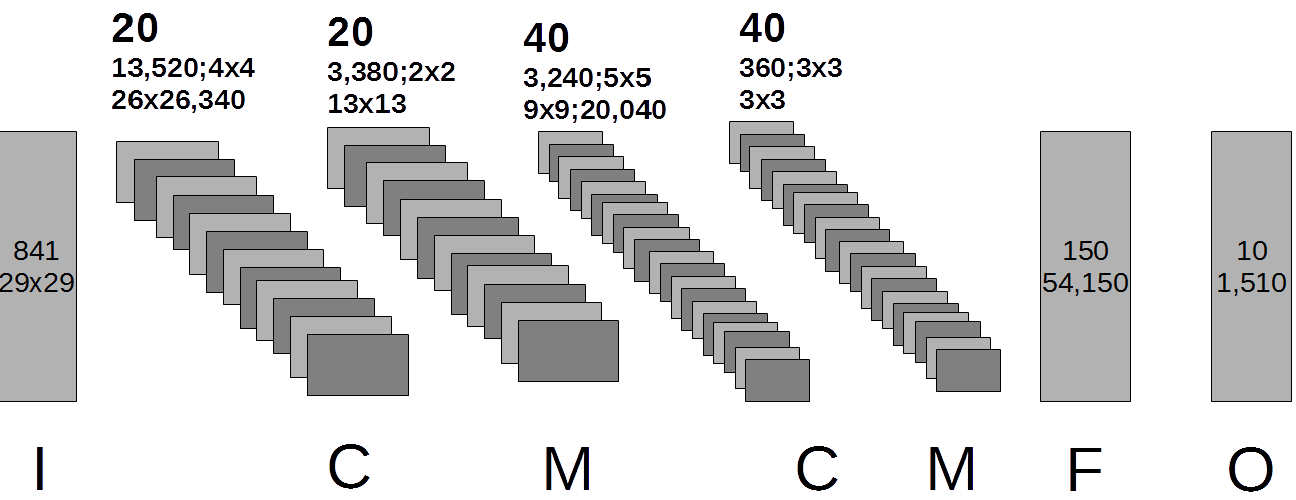}
       \label{fig:cnn-medium}}
	\hfill
    \subfloat[Large CNN: the last convolutional layer has 100 maps, 3,600 neurons, a 6x6 kernel, a map size of 6x6 and 216,100 weights.]
        {\includegraphics[width=\linewidth]{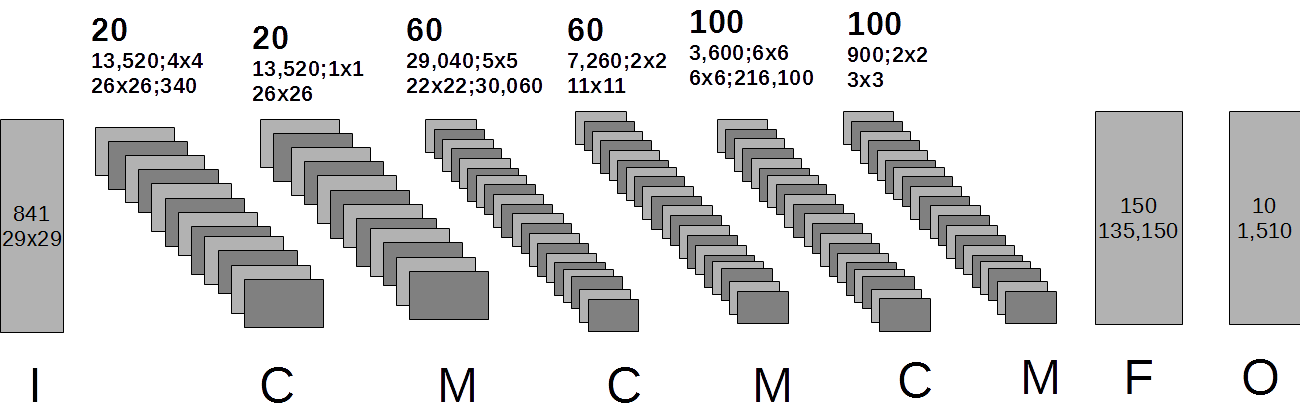}
        \label{fig:cnn-large}}
	\caption{CNN architectures used for experimental evaluation in this study. \emph{I} stands for input, \emph{C} for convolutional, \emph{M} for max-pooling, \emph{F} for fully connected and \emph{O} for output. The input layer has 841 neurons in a 29x29 grid. The output layer has 10 neurons.}
    \label{fig:cnn-architectures}
\end{figure}

DNNs can make predictions by forward propagating an input through the network. Forward propagation proceeds by performing calculations at each layer until reaching the output layer, which contains a vector representing the prediction. For example, in image classification problems, the output layer contains the prediction score that indicates the likelihood that an image belongs to a category \cite{agibansky,ng2011ufldl}. 
	
The forward propagation starts from a given input layer, then at each layer the activation for a neuron is activated using the equation $y^l_i = \sigma(x^l_i) + I^l_i$ where $y^l_i$ is the output value of neuron $i$ at layer $l$, $x^l_i$ is the input value of the same neuron, and $\sigma$ (sigmoid) is the activation function. $I^l_i$ is used for the input layer when there is no previous layer. The goal of the activation function is to return a normalized value (\textit{sigmoid} return [0,1] and \textit{tanh} is used in cases where the desired return values are [-1,1]). The input $x^l_i$ can be calculated as $x^l_i = \sum_j(w^{l}_{ji}y^{l-1}_j)$ where $w^{l}_{ji}$ denotes the weight between neuron $i$ in the current layer $l$, and $j$ in the previous layer, and $y^{l-1}_j$ the output of the $j$th neuron at the previous layer. This process is repeated until reaching the output layer. At the output layer, it is common to apply a soft max function, or similar, to squash the output vector and hence derive the prediction. 

Back-propagation is the process of propagating errors, i.e. the loss calculated as the deviation between the predicted and the desired output, backward in the network, by adjusting the weights at each layer. The error and partial derivatives $\delta^l_i$ are calculated at the output layer based on the predicted values from forward propagation and the labeled value (the correct value). At each layer, the relative error of each neuron is calculated and the weight parameters are updated based on how much the neuron participated in the faulty prediction. The expression $\delta E / \delta y^l_i = \sum{(w^l_{ij}  \delta E / \delta x^{l+1}_j})$ denotes that the partial derivative of neuron $i$ at the current layer $l$ is the sum of the derivatives of connected neurons at the next layer multiplied with the weights, assuming $w^l$ denotes the weights between the maps. Additionally, a decay is commonly used to control the impact of the updates, which is omitted in the above calculations. More concretely, the algorithm can be thought of as updating the layer\textquotesingle s weights based on "how much it was responsible for the errors in the output" \cite{agibansky,ng2011ufldl}.

A CNN is a multi-layer model constructed to learn various levels of representations where higher level representations are described based on the lower level ones \cite{schmidhuber2015deep}. It is a variant of deep neural network that introduces two new layer types: \emph{convolutional} and \emph{pooling} layers.
	
The \emph{convolutional layer} consists of several feature maps where neurons in each map connect to a grid of neurons in maps in the previous layer through overlapping kernels. The kernels are tiled to cover the whole input space. The approach is inspired by the receptive fields of the mammal visual cortex. All neurons of a map extract the same features from a map in the previous layer as they share the same set of weights. \emph{Pooling layers} intervene convolutional layers and have shown to lead to faster convergence. Each neuron in a pooling layer outputs the (maximum/average) value of a partition of neurons in the previous layer, and hence only activates if the underlying grid contains the sought feature. Besides from lowering the computational load, it also enables position invariance and down samples the input by a factor relative to the kernel size \cite{lecun1998gradient}.

LeNet-5 is an example of a Convolutional Neural Network. Each layer of convolution and pooling (that is a specific method of sub-sampling used in LeNet) comprise several feature maps. Neurons in the feature map cover different sub-fields of the neurons from the previous layer. All neurons in a map share the same weight parameters, therefore they extract the same features from different parts of the input from the previous layers. CNNs are commonly constructed similarly to the LeNet-5, beginning with an input layer, followed by several convolutional/pooling combinations, ending with a fully connected layer and an output layer \cite{lecun1998gradient}. 
	
In this study, the MNIST \cite{lecun2010mnist} dataset of handwritten digits is used. In total the MNIST data-set comprises 70000 images, 60000 of which are used for training/validation and the rest for testing. Figure \ref{fig:cnn-architectures} depicts three different CNN architectures that we use for evaluation: \emph{small, medium} and \emph{large}. There are various CNN implementations, such as, \emph{EbLearn} at New York University and \emph{Caffe} at Berkeley. As a basis for our work we selected a project developed by Cire{\c{s}}an \cite{Dan:code}, which targets the MNIST dataset of handwritten digits and has the possibility to dynamically configure the definition of layers, the activation function, and the connection types using a configuration file.

\section{Intel Many Integrated Core Architecture}
	\label{sec:xphi}
	
	\begin{figure}[t]
		\begin{center}
			\includegraphics[width=\linewidth]{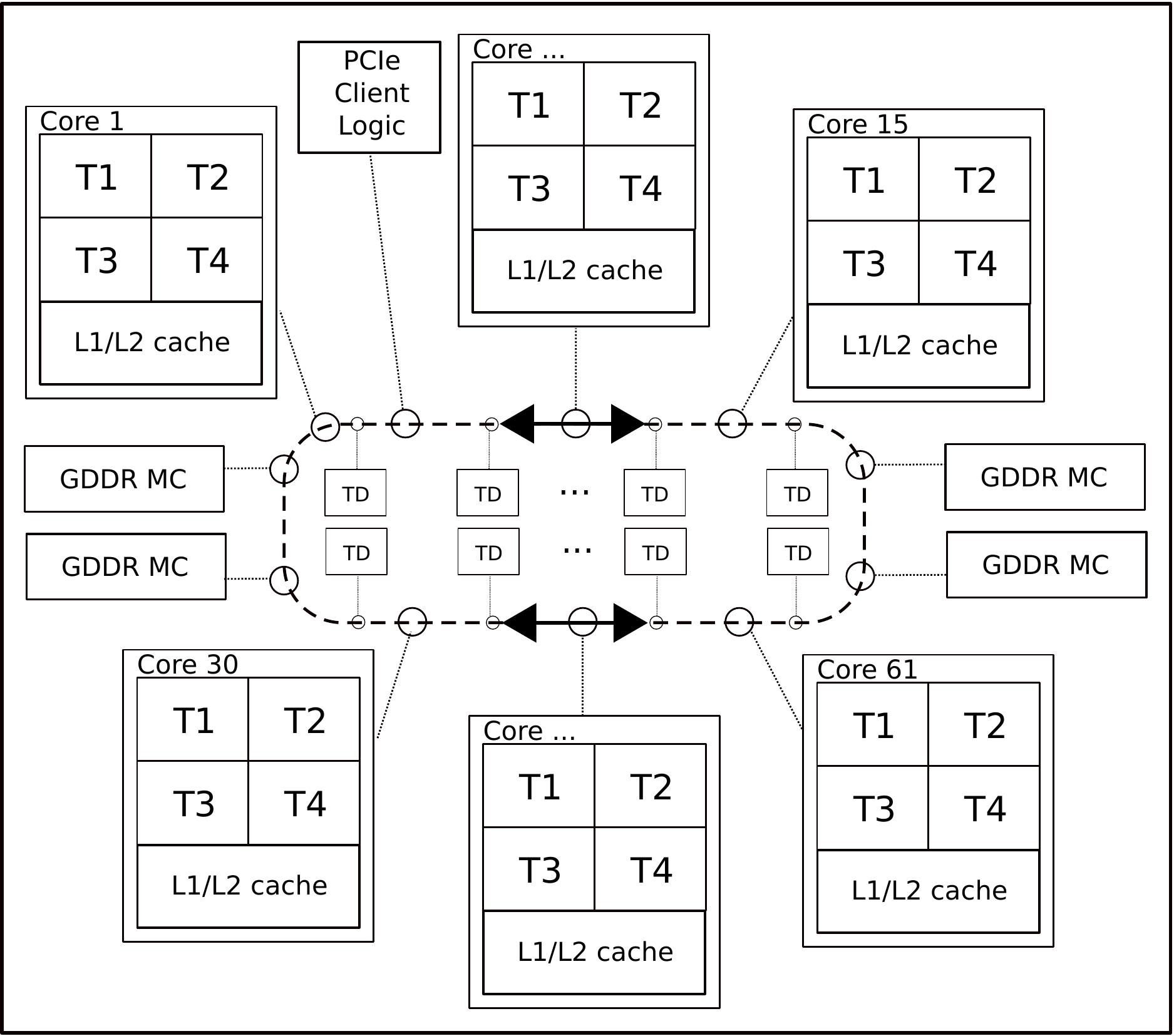}
		\end{center}
		\caption{An  example  of the Intel Many Integrated Core Architecture: Intel Xeon Phi.}
		\label{fig:emil-platform}
	\end{figure}
	
	Figure \ref{fig:emil-platform} depicts an overview of the Intel Xeon Phi (codenamed Knights Corner) architecture, which is an example of the Intel Many Integrated Core (MIC) Architecture. It is a many-core shared-memory Intel Xeon Phi processor, which runs a lightweight Linux operating system that offers the possibility to communicate with it over \emph{ssh}. The Intel Xeon Phi processor used in our study runs a $\mu$OS of version 2.6.38.8 and a software stack MPSS version 3.1.1. 
	
	The Intel Xeon Phi used in this study is of model 7120p, and facilitates 61 cores, each with a clock frequency of 1.2 GHz \cite{chrysos2012intel}. Each core can switch between four hardware threads in a round-robin manner, which amounts to a total of 244 threads per processor. Theoretically, the processor can deliver up to one teraFLOP/s of double precision performance, or two teraFLOP/s of single precision performance. Each core has its own L1 (32KB) and L2 (512KB) cache. The L2 cache is kept fully coherent by a global distributed tag-directory (TD). The cores are connected through a bidirectional ring bus interconnect, which forms a unified shared L2 cache of 30.5MB. In addition to the cores, there are 16 memory channels that in theory offer a maximum memory bandwidth of 352GB/s.

	Efficient usage of the available vector processing units of the Intel Xeon Phi is essential to fully utilize the performance of the processor \cite{TianSPGKMCP13}. Through the 512-bit wide SIMD registers it can perform 16 (16 wide $\times$ 32 bit) single-precision  or 8 (8 wide $\times$ 64 bit) double-precision operations per cycle. The Xeon Phi offers two programming models:
	
	\begin{enumerate}
		\item \emph{offload} - parts of the applications running on the host are offloaded to the Intel Xeon Phi processor
		\item \emph{native} - the code is compiled specifically for running natively on the Intel Xeon Phi processor. The code and all the required libraries should be transferred on the device. In this study, we use the native mode.
	\end{enumerate}

In this study, we use OpenMP \cite{Memeti:2017} for code implementation that exploits thread- and SIMD-parallelism available on the Intel Xeon Phi. The Intel Compiler 15.0.0 was used for native compilation of the application for the processor, whereas the $O3$ level was used for optimization. 

\section{Performance Modelling}
\label{sec:performance-model}

\begin{figure}[tb]
\center
\includegraphics[width=0.96\linewidth]{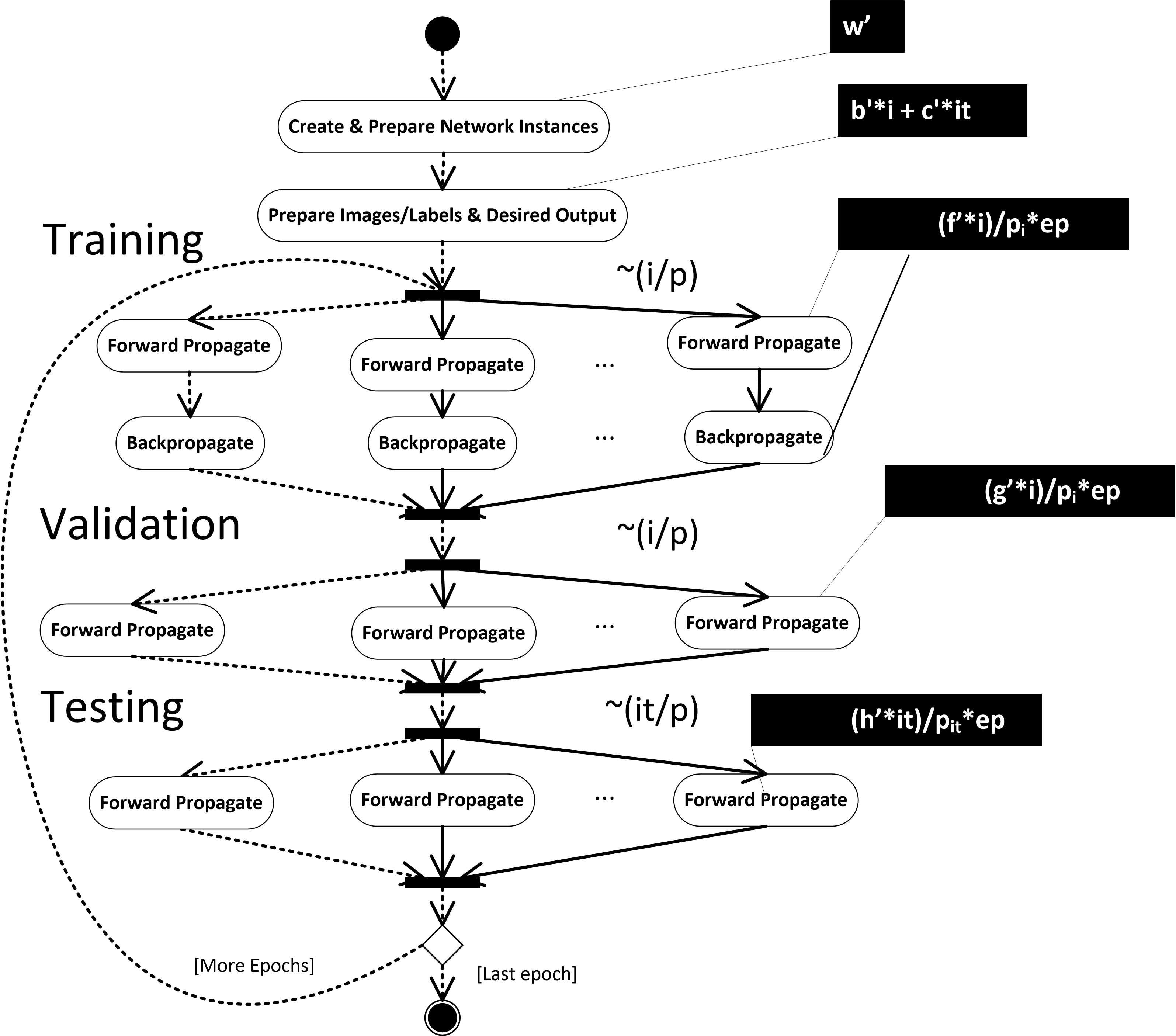}
\caption{An overview of our parallel deep leaning algorithm for Intel Xeon Phi. \textit{i} is the number of training or validation images, \textit{it} is the number of test images, \textit{ns} is the number of network instances, \textit{ep} is the number of epochs, \textit{p} is the number of processing units. w', b', c', f', g', h' indicate the work at various algorithmic steps.}
\label{fig:analysis_overview_a}
\end{figure}

A performance model \cite{perfmod,Fahringer:2004} enables us to reason about the behavior of an implementation in future execution contexts. Our performance model can predict the performance for numbers of threads that go beyond the number of hardware threads supported in the Intel Xeon Phi model that we used for evaluation. Additionally, it can predict the performance of different CNN architectures with various number of images and epochs. 

The input variables of our performance model $T(i,it,ep,p,s)$ are: the number of training or validation images (\textit{i}), the number of test images (\textit{it}), the number of network instances (\textit{ns}), the number of epochs (\textit{ep}), and the number of processing units (\textit{p}).

Figure {\ref{fig:analysis_overview_a}} depicts an overview of our parallel deep leaning algorithm for Intel Xeon Phi using call outs to denote the time complexity for different operations. Dashed lines denote the critical path through the algorithm. As each processing unit carries out equal amount of work, doing so in parallel reduces the overall computations required per worker, the shortest execution time depends on the slowest worker. Here, the creation of network instances is not parallelized. The span can be thought of as the sequential amount of work required to initialize images and labels, and other variables necessary, plus the maximum time for each network instances to carry out its intended amount of work in training, validation, and testing. If applying infinite number of processing units, what remains are the initial amount of work and the maximum time spent by each processing unit to process its chunk of the images.

The total execution time depends on several factors including: speed, number of processing units, communication costs (such as network latency), and memory contention. Of particular interest are contentions causing waiting times, including memory latencies and synchronization overhead. A time penalty referred to as $T_{mem}$ is added to the model to reflect memory and synchronization overhead. The contention is measured through an experimental approach by executing a small script on the Intel Xeon Phi processor for different thread counts, CNN weights and layers. The full set of variables is shown in Table \ref{tab:all_paramaters}.

\begin{table}[tb]
\center
\caption{Variables used in the performance model.}
\label{tab:all_paramaters}
\begin{tabular}{@{}ll@{}}
\toprule
Variable & Explanation\\ 
\midrule
\multicolumn{2}{c}{Parameters} \\
\cmidrule(lr){1-2}
\textit{p} & Number of processing units \\ 
\textit{i}&  Number of training/validation images\\ 
\textit{it} & Number of test images \\ 
\textit{ep} & Number of epochs \\ 
\midrule
\multicolumn{2}{c}{Constants - hardware dependent} \\
\cmidrule(lr){1-2}
\textit{CPI} & Best theoretical CPI/thread \\ 
\textit{s} & Speed of processing unit\\ 
\textit{OperationFactor} & Operation factor \\ 
\midrule
\multicolumn{2}{c}{Measured - hardware dependent} \\
\cmidrule(lr){1-2}
\textit{MemoryContention} & Memory contention \\ 
\textit{$T_{Fprop}$}+ & Forward propagation / image (ms)  \\ 
\textit{$T_{Bprop}$}+ & Back-propagation / image (ms) \\\textit{$T_{Prep}$}+ & Time for preparations \\  
\midrule 
\multicolumn{2}{c}{Calculated - hardware independent} \\
\cmidrule(lr){1-2}
\textit{FProp}* & \# FProp Operations / image \\ 
\textit{BProp}* & \# BProp Operations / image\\ 
\textit{Prep}* & \# Operations carried out for preparations \\
\midrule
\multicolumn{2}{l}{* The parameter is only used in prediction strategy (a)} \\
\multicolumn{2}{l}{+ The parameter is only used in prediction strategy (b)} \\
\bottomrule
\end{tabular}
\end{table}

\begin{table}[tb]
\center
\caption{Hardware independent parameters used in the performance model.}
\label{tab:model_parameters}
\begin{tabular}{@{}ll@{}}
\toprule
Term & Value \\ 
\midrule
\textit{Epochs (ep)} & 70 (small, medium), 15 (large) \\ 
\textit{Images (i)} & 60,000  \\ 
\textit{Images (it)} & 10,000  \\ 
\textit{Processing units/threads (p)} & 1 - 3,840\\
\midrule
\textit{FProp} & See \textit{table \ref{tab:forwardpropArch}} \\ 
\textit{BProp} & See \textit{table \ref{tab:backwardpropArch}} \\ 
\midrule
\multirow{3}{*}{\textit{Prep}} & Small:   $10^9$  \\ 
& Medium:  $10^{10}$   \\ 
& Large:    $10^{11}$   \\ 
\bottomrule
\end{tabular}
\end{table}

\begin{table}[tb]
\center
\caption{Hardware specific parameters used in the performance model.}
\label{tab:phi_parameters}
\footnotesize
\begin{tabular}{@{}ll@{}}
\toprule
Parameter & Intel Xeon Phi \\ 
\midrule
\textit{s} & 1.238 GHz  \\ 
\textit{Max processing units (p)} & 244 (240 used for prediction) \\ 
\midrule
\multirow{3}{*}{\textit{$T_{Fprop}$}(ms)}  & Small: 1.45  \\ 
& Medium: 12.55  \\ 
& Large: 148.88  \\ 
\midrule
\multirow{3}{*}{\textit{$T_{Bprop}$}(ms)} & Small: 5.3 \\ 
& Medium: 69.73 \\ 
& Large: 859.19  \\ 
\midrule
\multirow{3}{*}{\textit{$T_{Prep}$}(s)} & Small: 12.56 \\ 
& Medium: 12.7 \\ 
& Large: 13.5  \\ 
\midrule 
\textit{CPI} & 1-2 threads: 1; 3 threads: 1.5; 4 threads: 2  \\
\textit{MemoryContention} &  Table \textit{\ref{tab:phi_memory}} \\
\midrule
\multirow{3}{*}{\textit{OperationFactor}} & Small:  15  \\ 
& Medium:  15  \\ 
& Large:    15   \\ 
\bottomrule
\end{tabular}
\end{table}

\begin{table}[tb]
\center
\caption{Measured and predicted memory contention in seconds [s] for the Intel Xeon Phi.}
\label{tab:phi_memory}
\begin{tabular}{@{}rlll@{}}
\toprule
\# Threads & Small CNN & Medium CNN & Large CNN\\ 
\midrule 
 1 & $7.10*10^{-6}$ & $1.56*10^{-4}$ & $8.83*10^{-4}$\\
 15 & $6.40*10^{-4}$ & $2.00*10^{-3}$ & $8.75*10^{-3}$\\ 
 30 & $1.36*10^{-3}$ & $3.97*10^{-3}$ &  $1.67*10^{-2}$ \\ 
 60 & $3.07*10^{-3}$ & $8.03*10^{-3}$ &  $3.22*10^{-2}$ \\ 
 120 & $6.76*10^{-3}$ & $1.65*10^{-2}$ & $6.74*10^{-2}$\\ 
 180 & $9.95*10^{-3}$ & $2.50*10^{-2}$ & $1.00*10^{-1}$\\ 
 240 & $1.40*10^{-2}$ & $3.83*10^{-2}$ &  $1.38*10^{-1}$\\ 
 480* & $2.78*10^{-2}$ & $7.31*10^{-2}$ &	$2.73*10^{-1}$ \\
 960* & $5.60*10^{-2}$ & $1.47*10^{-1}$ & $5.46*10^{-1}$ \\
 1,920* & $1.12*10^{-1}$ & $2.95*10^{-1}$ & $1.09$ \\
 3,840* & $2.25*10^{-1}$ & $5.91*10^{-1}$ &  $2.19$ \\
\midrule
\multicolumn{4}{l}{* Predicted memory contention}  \\
\bottomrule
\end{tabular}
\end{table}

We define memory overhead as, $T_{mem}(ep,i,p) = (MemoryContention * ep *i)/p$ where $MemoryContention$ is the measured memory contention when $p$ threads are competing for I/O concurrently. The measured and predicted values for memory contention are depicted in Table \ref{tab:phi_memory}. In Table \ref{tab:all_paramaters} parameters used in the performance model are depicted; some parameters are hardware dependent and others independent of the underlying hardware. Each parameter is either measured or calculated. Table \ref{tab:model_parameters} shows parameters that are independent of the hardware, and Table \ref{tab:phi_parameters} shows the parameters that are specific for the Intel Xeon Phi.

We follow two strategies for performance modelling: 
\begin{itemize}
    \item Strategy (a) minimizes the use of measurements for estimating parameter values of the performance model. Only memory contention is estimated using measurements. The performance model for strategy (a) is depicted in Table~\ref{tab:model_a}.
    \item Strategy (b) applies the measurements to estimation of the sequential work, the forward- and back-propagation. The performance model for strategy (b) is depicted in Table~\ref{tab:model_b}.
\end{itemize}

\begin{table}[ht]
\centering
\caption{Performance model for strategy (a).}
\label{tab:model_a}
\begin{tabular}{l|l}
 & $T(i,it,ep,p,s)$ \\
 & $= T_{comp}(i,it,ep,p,s) + T_{mem}(ep,i,p)$ \\
preparation work & $= \Bigg(\dfrac{Prep + 4*i +2*it+ 10*ep}{s}$ \\
training & $+ \Bigg(\bigg(\Big(\dfrac{FProp+BProp}{s}\Big)*\dfrac{i}{p_i}*ep\bigg)$ \\
validation  & $+ \bigg(\Big(\dfrac{FProp}{s}\Big)*\dfrac{i}{p_i}*ep\bigg)$\\
testing & $+ \bigg(\Big(\dfrac{FProp}{s}\Big)*\dfrac{it}{p_{it}}*ep\big)\Bigg)$\\
calculation penalty & $* CPI\Bigg) * OperationFactor$ \\
memory overhead & $+ \dfrac{MemoryContention*i*ep}{p}$ \\
\end{tabular}
\end{table}

Please note that the constants are approximations, they are relative to each other, and yet far from precise. $Prep$ is different for each CNN architecture ($10^9, 10^{10}$ and $10^{11}$ for small, medium and large architecture respectively) and denotes the number of operations required to create network instances, prepare weights, etc. The $OperationFactor$ is adjusted to closely match the measured value for 15 threads, and mitigate the approximations done for instructions in the first place, at the same time account for vectorization.

\begin{table}[ht]
\centering
\caption{Performance model for strategy (b).}
\label{tab:model_b}
 \begin{tabular}{l|l}
  & $T(i,it,ep,p)$ \\
  & $= T_{comp}(i,it,ep,p) + T_{mem}(ep,i,p)$ \\
 preparation work & $= T_{prep}$ \\
 training & $+ \Bigg(\Big(\big(T_{FProp}+T_{Bprop}\big)*\dfrac{i}{p_i}*ep\Big)$ \\
 validation  & $+ \Big(T_{FProp}*\dfrac{i}{p_i}*ep\Big)$\\
 testing & $+ \Big(T_{FProp}*\dfrac{it}{p_{it}}*ep \Big) \Bigg)$\\
 calculation penalty & $* CPI$ \\
 memory overhead & $+ \dfrac{MemoryContention*i*ep}{p}$ \\
 \end{tabular}
 \end{table}

$T_{prep}$ is the measured time it takes to prepare the training (small 12.56 seconds, medium 12.7 seconds, and large 13.5 seconds); $T_{FProp}$ and $T_{BProp}$ indicate the required time to forward- and back-propagate one image through the network. When one hardware thread is available per core, then one instruction per cycle can be assumed. For four threads per core, only 0.5 instructions per cycle can be assumed per thread; each thread gets to execute two instructions every fourth cycle ($CPI$ of 2). The speed $s$ is defined in Table \ref{tab:phi_parameters}. $FProp$ and $BProp$ are placeholders for the actual number of operations shown in Table \ref{tab:forwardpropArch} and Table \ref{tab:backwardpropArch} respectively. 

\begin{table}[tb]
\center
\caption{$FProp$: number of operations when forward propagating one image for small, medium and large CNN architectures.}
\label{tab:forwardpropArch}
\begin{tabular}{lrrrrr}
\toprule 
 & Max Pool. & Fully Con. & Convolution & Total & Ratio\\ 
\midrule 
\textit{Small} &  7k & 5k &  46k & 58k & - \\  
\textit{Medium} &  29k &  56k &  474k & 559k & 9.64 \\  
\textit{Large} &  99k &  137k &  5,113k & 5,349k & 9.57 \\ 
\bottomrule
\end{tabular} 
\end{table}

\begin{table}[tb]
\center
\caption{$BProp$: number of operations when back propagating one image for small, medium and large CNN architectures.}
\label{tab:backwardpropArch}
\begin{tabular}{lrrrrr}
\toprule 
 & Max Pool. & Fully Con. & Convolution & Total & Ratio \\ 
\midrule 
\textit{Small} & 2k & 10k  & 512k & 524k & - \\  
\textit{Medium} & 4k & 112k & 6,003k & 6,119k & 11.68 \\  
\textit{Large} & 8k & 274k &  72,896k & 73,178k & 11.96 \\
\bottomrule 
\end{tabular} 
\end{table}

\section{Evaluation of Performance Model}
\label{sec:evaluation}
	
In this section, we compare the predicted and measured execution times for various numbers of threads and CNN architectures. The execution time is the total time the program runs, excluding the time required to initialize the network instances and images. To evaluate our approach we use an Intel Xeon Phi 7120P accelerator that comprises 61 cores that run at 1.2 GHz. We use 1, 15, 30, 60, 120, 180, and 240 threads of the Intel Xeon Phi processor. Each thread is responsible for one network instance. In the figures, we use the following notations: \emph{Par} refers to the parallel version, and \emph{T} denotes threads, for instance, \emph{Phi Par. 120 T} is the parallel version that is executed by 120 threads on the Intel Xeon Phi.

\textbf{Result 1:} \emph{The predicted execution times obtained from the performance model match well the measured execution times.}
	
\begin{figure}[bt]
		\center
		\includegraphics[width=\linewidth]{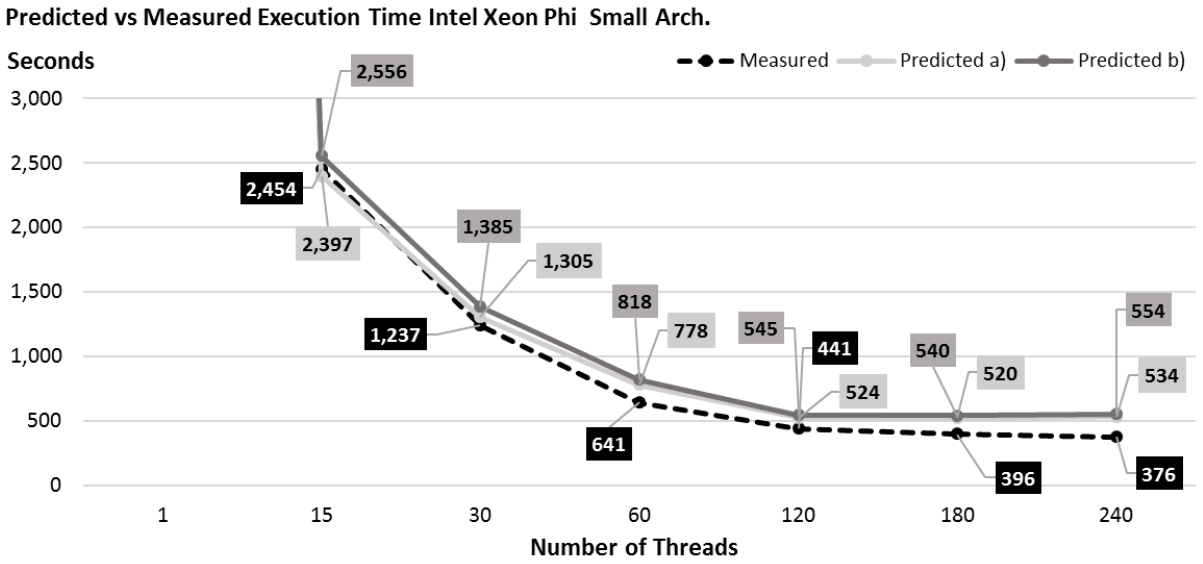}
		\caption{Comparing predicted execution times with measured execution times on Intel Xeon Phi for the small CNN architecture.}
		\label{fig:pred-model-small}
	\end{figure}

\begin{figure}[bt]
		\center
		\includegraphics[width=\linewidth]{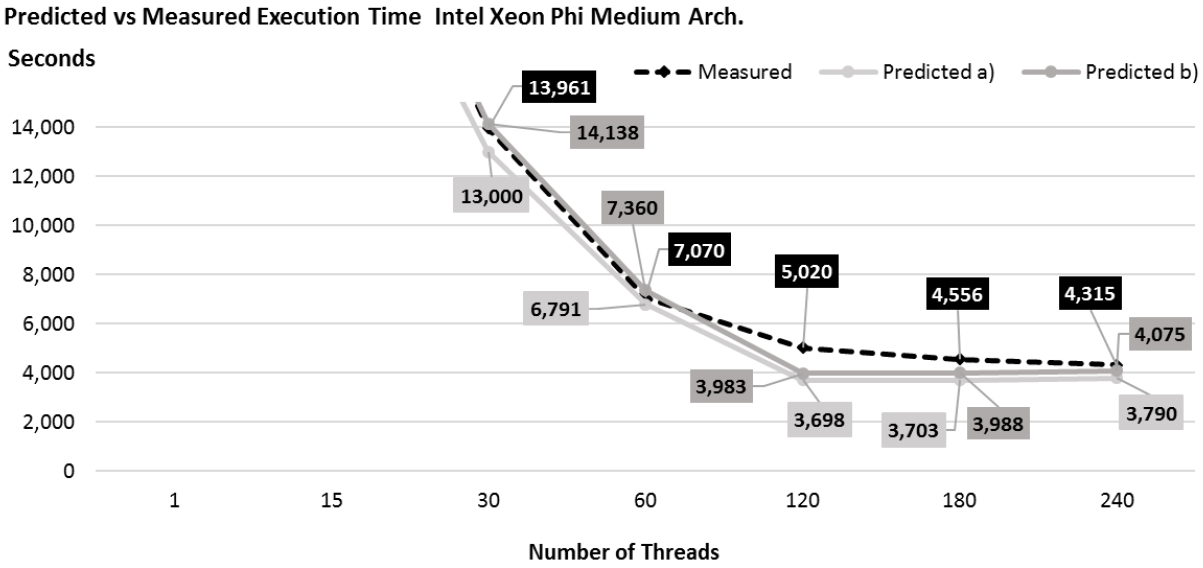}
		\caption{Comparing predicted execution times with measured execution times on Intel Xeon Phi for the medium CNN architecture.}
		\label{fig:pred-model-medium}
	\end{figure}

\begin{figure}[bt]
		\center
		\includegraphics[width=\linewidth]{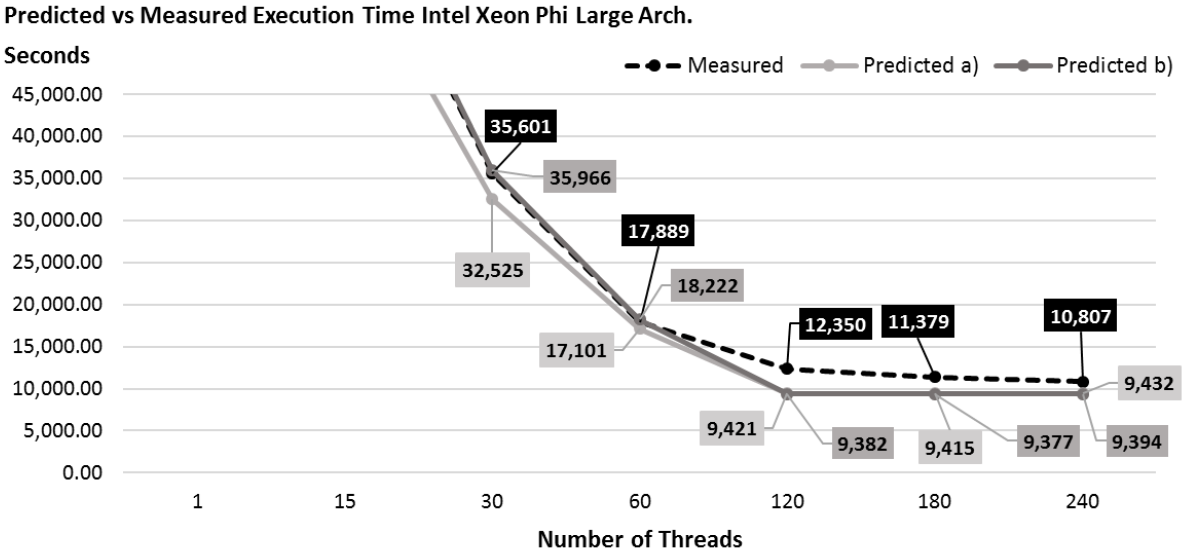}
		\caption{Comparing predicted execution times with measured execution times on Intel Xeon Phi for the large CNN architecture.}
		\label{fig:pred-model-large}
	\end{figure}

Figures \ref{fig:pred-model-small}, \ref{fig:pred-model-medium}, and \ref{fig:pred-model-large} depict the predicted and measured execution times for small, medium and large CNN architecture. For the small network (Figure \ref{fig:pred-model-small}), the predictions are close to the measured values with a slight deviation at the end. The prediction model seems to over-estimate the execution time with a small factor. 
	
For the medium architecture (Figure \ref{fig:pred-model-medium}) the prediction follow the measured values closely, although it underestimates the execution time slightly. At 120 threads, the measured and predicted values start to deviate, which are recovered at 240 threads.
	
The large CNN architecture (Figure \ref{fig:pred-model-large}) yields similar performance results as the medium CNN architecture. We may observe that the measured values are slightly higher than the predictions, however, the predictions follow the measured values. For 120 threads there is a deviation between the measured and predicted value, which is then improved for 240 threads. While the predicted execution time increases between 120 and 240 threads, the measured execution time decreases. This is most probably due to the CPI factor that is added when 3 or more threads are present on the same core. 

We use the expression $\Delta = (\abs{T_\mu-T_\psi}/{T_\psi})100\%$ to calculate the prediction accuracy of our performance model, where $T_\mu$ is the measured and $T_\psi$ is the predicted value. The average prediction accuracy for strategies (a) and (b) and various CNN architectures is shown in Table \ref{tab:error_in_prediction_percentage}. We may observe, that model (a) is more accurate for the small CNN, whereas the model (b) is better for medium and large CNNs. 
	
\begin{table}[tb]
\center
\caption{Average accuracy $\Delta$ of performance model for prediction strategies (a) and (b) and all considered CNN architectures.}
\label{tab:error_in_prediction_percentage}
\begin{tabular}{@{}ccccccc@{}}
\toprule
& \multicolumn{2}{c}{Small CNN} & \multicolumn{2}{c}{Medium CNN} & \multicolumn{2}{c}{Large CNN} \\ 
& a &  b &  a &  b &  a &  b 
\\
\midrule
 & 14.57\%	& 16.35\%& 	14.76\%	& 7.48\%	& 15.36\%& 	10.22\% \\
 \bottomrule
\end{tabular}
\end{table}
	
\textbf{Result 2:} \textit{The performance model results indicate that CNN training on Intel Xeon Phi scales well up to several thousands of threads.}
	
We used the prediction model to predict the execution times for 480, 960, 1920, and 3840 threads for different CNN architectures, using the same parameters. The results in Table \ref{tab:pred-model-results} show that if \textit{3,840} threads were available, the small network should take about \textit{4.6} minutes to train, the medium \textit{14.5} minutes and the large \textit{36.8} minutes. The predictions for the large CNN architecture are not as well aligned when increasing to larger thread counts as for small and medium. 
	
\begin{table}[t]
\center
\caption{Predicted execution times in minutes for 480, 960, 1,920 and 3,840 images using the performance models (a) and (b).}
\label{tab:pred-model-results}
    \begin{tabular}{@{}rcccccc@{}}
    \toprule
    & \multicolumn{2}{c}{Small} & \multicolumn{2}{c}{Medium} & \multicolumn{2}{c}{Large} \\ 
    & a &  b &  a &  b &  a &  b 
    \\
    \midrule
    480  & 6.6 & 6.7 & 36.8 & 39.1 & 92.9 & 82.6 \\ 
    960  & 5.4 & 5.5 & 23.9 & 25.1 & 60.8 & 45.7 \\
    1,920 & 4.9 & 4.9 & 17.4 & 18.0 & 44.8 & 27.2 \\
    3,840 & 4.6 & 4.6 & 14.2 & 14.5 & 36.8 & 18.0 \\ 
    \bottomrule
\end{tabular}
\end{table}
	
	Additionally, we evaluated the execution time for varying image counts, and epochs, for 240 and 480 threads for the small CNN architecture. As can be seen in Table \ref{tab:futurepredictions} doubling the number of images or epochs, approximately doubles the execution time. However, doubling the number of threads does not reduce the execution time in half.
	
	\begin{table}[t]
		\centering
		\caption{The execution times in minutes when scaling epochs and images for 240 and 480 threads using the performance model (a) on the small CNN architecture.}
		\begin{threeparttable}
			\begin{tabular}{r r r r r r r r}
				\toprule
				& & \multicolumn{3}{c}{240 Threads} 
				& \multicolumn{3}{c}{480 Threads} 
				\\ \midrule
				\multicolumn{2}{c}{Images}             
				& \multicolumn{3}{c}{Epochs}      
				& \multicolumn{3}{c}{Epochs}      \\ 
				\multicolumn{1}{c}{$i$\tnote{1}} & \multicolumn{1}{c}{$it$\tnote{2}} & 
				70    & 140   & 280        & 70    & 140   & 280        
				\\ \midrule
				\multicolumn{1}{r}{60k} & 10k                     
				& 8.9   & 17.6  & 35.0      & 6.6   & 12.9  & 25.6    
				\\
				\multicolumn{1}{r}{120k} & 20k                     
				& 17.6  & 35.0  & 69.7     & 12.9  & 25.6  & 51.1     
				\\
				\multicolumn{1}{r}{240k} & 40k                     
				& 35.0  & 69.7  & 139.3    & 25.6  & 51.1  & 101.9    
				\\ \bottomrule
			\end{tabular}
			\begin{tablenotes}
				{ \footnotesize 
					\item[1] Number of images in the training/validation set
					\item[2] Number of images in the test set
				}
			\end{tablenotes}
		\end{threeparttable}
		\label{tab:futurepredictions}
	\end{table}

\section{Related work}
\label{sec:rw}

In this section, we discuss related work with respect to performance modeling of deep learning. 

Yan et al. \cite{yan2015} focus on performance modeling and optimization of deep learning on distributed systems. The authors use analytical performance modeling techniques to explore the configuration space and find optimal system configurations to minimize the iteration time over the training data. According to the authors, the error rates of under 25\% allow them to identify and distinguish good combination of system parameters from the not so good ones.

Oyama et al. \cite{oyama2016} propose a performance prediction model for an asynchronous stochastic gradient descent deep learning system. The proposed approach considers the probability distribution of mini-batch sizes and staleness (that is, the number of updates done within one gradient computation). The authors report model accuracy of 81-95\% for various mini-batch sizes. Similar to our work, the authors use the prediction model to evaluate the scalability of deep learning for upcoming hardware architectures.

Paleo, a performance model proposed by Qi et al. \cite{qi2017paleo}, can efficiently predict a combination of the network architecture, hardware and software choices, parallelization strategies, and communication schemes to model the expected performance and scalability of training deep neural networks.

Song et al. \cite{song2017}, in contrast, focus on the different requirements that the end-users need to perform various prediction tasks. They propose an approach that combines offline compilation (to select optimal batch-sizes) and run-time management (to identify and schedule the fastest kernels, and partition the available resources accordingly). The authors use analytical models to predict the optimal resources of a GPU (such as streaming multiprocessors) to use in each layer, and predict the processing time of a given layer.

Shi et al. \cite{shi:2018} use performance modelling to evaluate various distributed deep learning frameworks (such as, Caffe-MPI or TensorFlow) on GPU accelerated computing systems. Authors observe performance gaps between deep learning implementations under study and identify methods that require further optimization. 

Yufei et al. \cite{yufei2019} propose a performance model for prediction of throughput on FPGAs, which is used to identify and explore optimal design choices during the design phase. The authors focus on modeling the DRAM access, latency, and on-chip buffer access. The validation results show that estimations derived from the model closely match (within 3\%) the actual test results executed on Arria 10 and Stratix 10 FPGAs.

In contrast to the related work, we focus on performance modeling of training deep convolutional neural networks on the Intel Xeon Phi many-core processor. In our previous work \cite{Memeti:2018b} we used machine learning for performance prediction of DNA sequence analysis \cite{Memeti:2015,Memeti:2016} on Intel Xeon Phi .  

\section{Summary}
\label{sec:summary}

Deep learning is essential for solving complex problems in many domains including, self-driving cars, object recognition, natural language processing, speech recognition, and language translation. In this paper, we have described an approach for performance modeling of training convolutional neural networks on the Intel Xeon Phi many core processor. We developed two parameterized performance models based on the theoretical code analysis. For the development of the first performance model, we minimally used measurements for estimating parameter values of the performance model; only memory contention was estimated using measurements. For the second performance model, we used measurements for estimation of the sequential work, and the forward- and back-propagation. We used three different convolutional neural network architectures for evaluation of performance prediction accuracy of the developed models. The average deviation of predicted from the measured performance over all measured thread counts and various neural network architectures was about 15\% for the first model and 11\% for second model.

Future work will develop performance models of deep learning on large-scale parallel computing systems that comprise multiple nodes with many-core processors.



\end{document}